\documentclass[a4paper,twoside,10pt]{article}

\input{jgrg19.sty}
\begin{document}
%
\pagestyle{fancy}
\fancyhead{}
  \fancyhead[RO,LE]{\thepage}
  \fancyhead[LO]{T. Matsuda}                  
  \fancyhead[RE]{Warm directions}    
\rfoot{}
\cfoot{}
\lfoot{}
\label{P18}    
\title{Inflation, modulation and baryogenesis with warm
directions}
%
\author{%
  Tomohiro Matsuda\footnote{Email address: matsuda@sit.ac.jp}}
%
\address{
  Department of Physics, Saitama Institute of Technology, Saitama, 
369-0293
}

\abstract{
There are many flat directions in SUSY models, which may dissipate their
energy and source the radiation background during inflation. 
However, the only possibility that has been studied in this direction
is warm inflation,
which uses ``warm'' (or ``dissipative'' if we consider more modest
situation) direction as the inflaton. 
In this talk we discuss other significant possibilities of such
directions which are dissipative and may or may not be ``warm''.
Affleck-Dine (AD) mechanism and other cosmological scenarios 
are discussed in the light of ``dissipative field'', 
instead of using the conventional light field with mass protection.
We sometimes consider Morikawa-Sasaki coefficient for the non-thermal
background, which is important because the dissipation calculated for a
naive thermal background with $T\rightarrow 0$ is not enough to discuss
the dissipation with the non-thermal background. (This is a small
extended version of the proceedings for the JGRG19.)} 



\section{Dissipative directions for particle cosmology}
Even in a non-thermal background, dissipation is generic for 
realistic field motion $\dot{\phi}\ne 0$ that leads to a coherent
excitation of a heavy intermediate field $\chi$ that decays into light
fermions $\psi_d$.
Here we consider a typical interaction given by
\begin{equation}
{\cal L}_{int}=-\frac{1}{2}g^2 \phi^2\chi^2 -h\chi\bar{\psi}\psi,
\end{equation}
which leads to efficient decay of the intermediate field ($m_\chi
\propto \phi$) with the decay rate 
\begin{equation}
\Gamma_{\chi}\simeq \frac{N_\psi}{8\pi} h^2 m_\chi
\simeq \frac{N_\psi}{8\pi} h^2 g \phi,
\end{equation}
where $N_\psi$ is the number of the light fermions.
The dissipation coefficient is given by \cite{gamma-MS, gamma-MS-new,
bose-d} 
\begin{equation}
\Upsilon \sim N_\chi 
\frac{\sqrt{2}g^3  N_\psi h^2\phi}{8^3\pi^2},
\end{equation}
which is proportional to $\Gamma_\chi$.
In figure 1, we show a schematic picture for the dissipative motion.
\begin{figure}[h]
 \begin{center}
\begin{picture}(100,118)(100,160)
\resizebox{8cm}{!}{\includegraphics{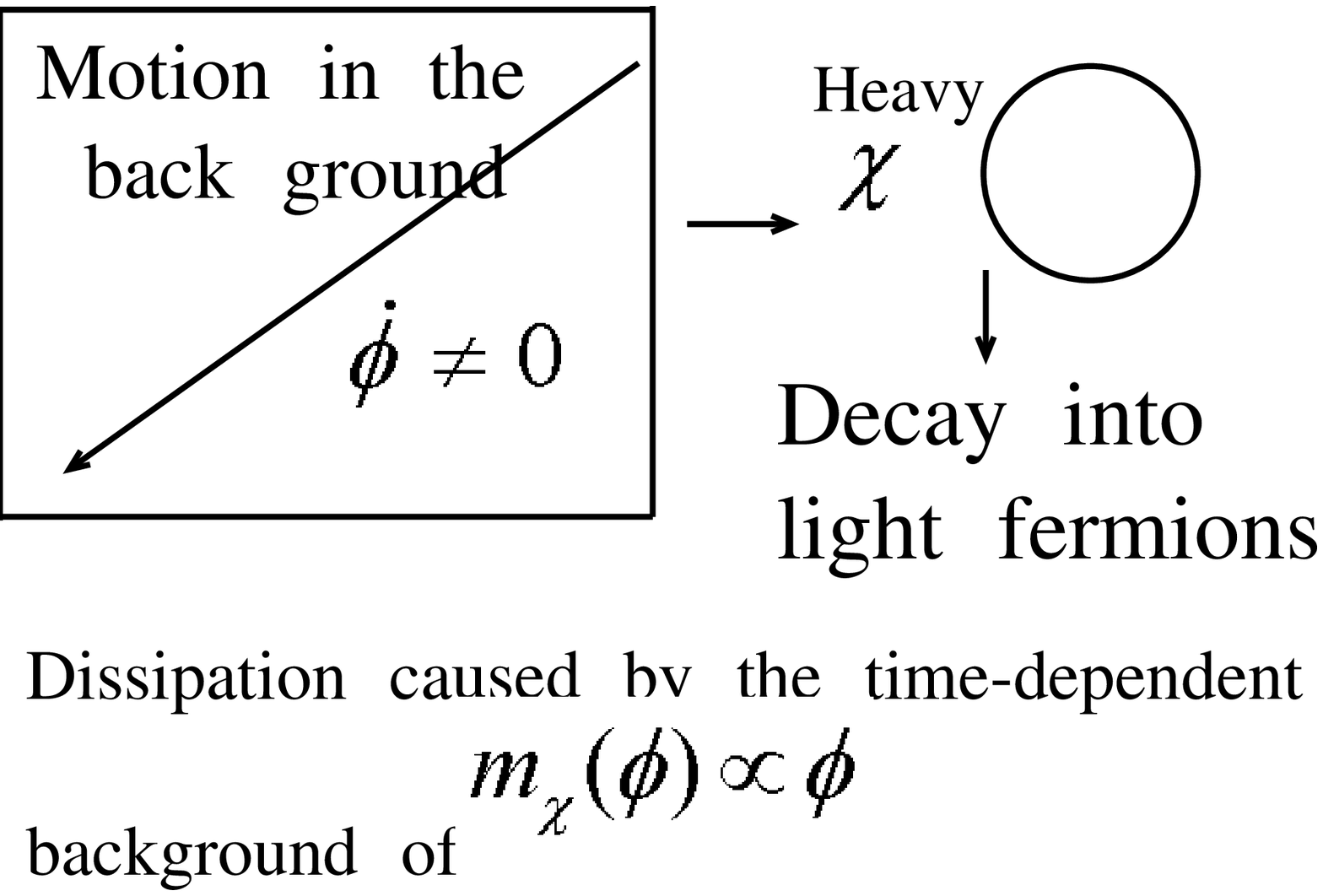}} 
\end{picture}
\caption{``Energy loss'' is caused by the correction that
has the imaginary part proportional to $\Gamma_\chi$.}
 \end{center}
\end{figure}

The strength of the ``friction'' caused by the dissipation is measured
by the rate $r_\Upsilon\equiv \frac{\Upsilon}{3H}$.
Then the field equation for the dissipative motion is given by
\begin{equation}
\ddot{\phi}+3H(1+r_\Upsilon)\dot{\phi}+V_\phi=0,
\end{equation}
where the subscript denotes the derivative with respect to the field.
The effective slow-roll parameters are suppressed when 
$r_\Upsilon$ is large (``strongly dissipating'');
\begin{equation}
\epsilon_{eff} =  \frac{\epsilon}{(1+r_\Upsilon)^2},\,\,\,\,
\,\,\,\,\,\, \eta_{eff} =  \frac{\eta}{(1+r_\Upsilon)^2}.
\end{equation}
Our claim is very simple.
{\bf Dissipative motion is a generic phenomenon, which must be considered
not only for the specific inflation model but also for other generic
cosmological scenarios.} 
A {\bf modest} assumption is that the background is {\bf not} thermal,
because ``warm'' background is not essential for the dissipative motion.
Of course dissipation may be more significant when the background is
thermal, but the required conditions for the thermalization are
sometimes very severe.
We thus consider non-thermal background during inflation because thermal
conditions may spoil the generality of the dissipative scenario.
On the other hand, thermal background is natural after reheating.
We thus consider thermal background for the field motion after
reheating.
Assuming a thermal background, $\Upsilon$ can be given by
\begin{equation}
\Upsilon \propto \frac{T^n}{\phi^{n-1}},
\end{equation}
where $n=1$ for high-temperature SUSY and $m=3$ for low-temperature
 SUSY.
\subsection{Natural dissipation in SUSY hybrid inflation}
The first example\cite{matsuda-2009k}
 is SUSY hybrid inflation, for which we will argue that
the conventional interaction 
${\cal L}_I \sim -\frac{1}{2}g^2\phi^2\chi^2$
between the inflaton $\phi$ and the trigger
field $\chi$ may cause significant dissipation that leads to slow-roll
inflation.
Namely, $O(H)$ correction from the supergravity may not spoil
slow-roll in SUSY hybrid inflation.

The key in this scenario is the gravitational decay $\chi\rightarrow
2\psi_{3/2}$ of the intermediate (trigger) field, which leads to an
inevitable decay rate 
\begin{equation}
\Gamma_{\chi\rightarrow 2\psi_{3/2}}\simeq \frac{m_\chi^3}{M_p^2}
\sim  \frac{g^3 \phi^3}{M_p^2},
\end{equation}
which is larger than the Hubble parameter $H$ when 
$\phi > (HM_p^2)^{1/3}/g$.
The dissipation coefficient of the inflaton mediated by the heavy
 trigger field $\chi$ is 
\begin{equation}
\Upsilon \sim 10^{-2}\left(\frac{m_\chi}{M_p}\right)^2 \phi,
\end{equation}
which gives the minimal value of $\Upsilon$ caused by
the least channel of the gravitational decay.
Surprisingly, $r_\Upsilon\gg 1$ is generic 
 for the chaotic initial condition $\phi_{ini} \sim M_p$.
See Fig.2.
\begin{figure}[h]
 \begin{center}
\begin{picture}(105,155)(250,180)
\resizebox{22cm}{!}{\includegraphics{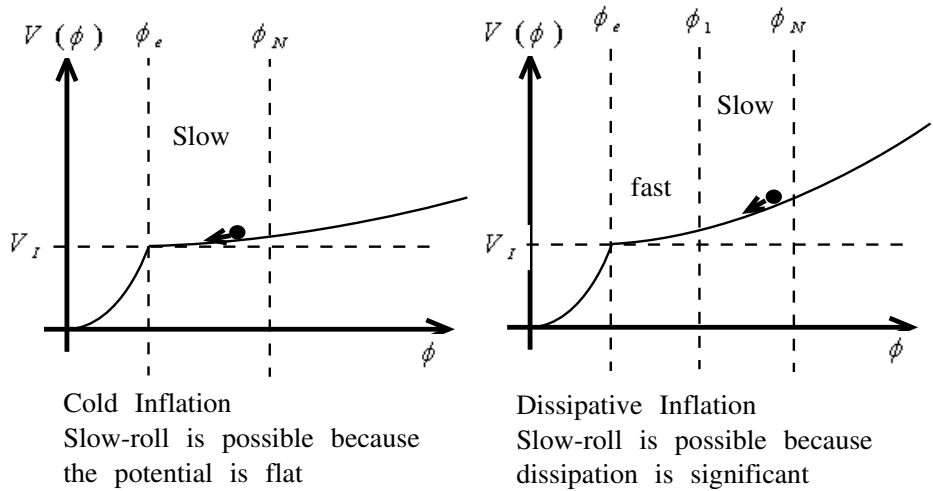}} 
\end{picture}
\caption{Left : Conventional cold inflation with $m_\phi\ll H$. 
Right : Gravitational decay of the trigger field causes
  significant dissipation and slow-roll for $\phi_1<\phi$ even if
  $m_\phi \ge H$. }
 \end{center}
\end{figure}
\subsection{Dissipative Affleck-Dine field}
Considering typical situation for the dissipative motion of the
AD field, it is not appropriate to disregard thermal background
$T\ne 0$. 
However, {\bf in contrast to warm inflation} \cite{warm-inf,warm-inf2}, 
the thermal background is
not always due to the dissipation caused by the field motion.
Namely, the dissipation coefficient $\Upsilon(\phi,T)$ of the
AD field may depend on the environment, temperature of the
Universe, which depends on cosmological events other than the
Affleck-Dine baryogenesis. 
Typically, MSSM directions couple to heavy directions that can
decay into light fermions. 
Therefore, non-thermal dissipation would be significant at large
distance, and thermal dissipation may be significant during a period
depending on the field interaction and the temperature of the Universe.
In any case, it is very important to consider dissipation of
the AD field before the AD baryogenesis.
If the dissipation is large enough to ensure the slow-roll,
the AD-field is ``trapped'' at $\phi_{AD}\ne 0$.
{\bf The situation is in contrast with the conventional scenario, in
which the ``flip'' of the potential is responsible for $\phi_{AD}\ne
0$.}
If the AD field is trapped due to dissipation, the time when
oscillation begins is not determined by the usual condition 
$m_\phi\sim H$.

The situation related to dissipation of the AD field 
can be summarized as follows.

1) The time when oscillation starts may be different from the
   conventional (non-dissipating) scenario.  

2) The amplitude of the AD-field oscillation may be different.

These differences are expected to lead to crucial discrepancies
in the results.

\subsection{Dissipation before preheating}
Above in section 1.1, we considered hybrid inflation in which the
typical interaction 
(inflaton-trigger field interaction) leads to inevitable dissipation.
A similar argument may apply to preheating scenario \cite{preheating,
preheating2, preheating3}
 in which the typical interaction (oscillating field-preheat field
 interaction) 
may lead to significant dissipation and slow-roll before the 
oscillation.
Here we consider ``instant preheating'' scenario for simplicity, in
which instant decay after preheating is assumed.
Significant dissipation before the onset of oscillation
typically leads to a {\bf delay of the oscillation}. 
To show explicitly the dissipative effect,
we consider a potential with a mass 
\begin{equation}
V(\phi)=\frac{1}{2}m^2\phi^2,
\end{equation}
and the dissipation based on the non-thermal background.
The interaction with the preheat field is given by
\begin{equation}
{\cal L}_{int}\simeq \frac{g_{PR}^2}{2} \phi^2 \chi^2,
\end{equation}
where {\bf preheat field} $\chi$ plays
the role of the {\bf intermediate} field for the dissipation.
The decay into light fermions, which is needed for the preheating
scenario followed by the instant decay, is induced by the term
\begin{equation}
{\cal L}_{\psi\chi}=h \chi\bar{\psi}\psi.
\end{equation}
Obviously, for $g\sim h\sim O(1)$, the dissipative process
$\phi\rightarrow \chi \rightarrow \psi$ is efficient for the model of
instant preheating.
In fact, based on the non-thermal dissipation, it is very easy to show
the slow-roll conditions that delays the preheating and reduces the
amplitude of the oscillation.
The dissipation coefficient is 
\begin{equation}
\Upsilon \simeq 10^{-2}g_{PR}^2 h^2 m_\varphi  
\simeq 10^{-2}g_{PR}^3 h^2 \phi.
\end{equation}
Then the effective slow-roll conditions are 
\begin{eqnarray}
\epsilon_w &\simeq& M_p^2 
\left(\frac{m^2 \phi_I}{V}\right)^2 
\frac{1}{(1+r)^2}
\sim  
\left(\frac{m}{10^{-2}\times  \phi_I }\frac{1}{g_{PR}^3 h^2}\right)^2 < 1
\\
\eta_w &\simeq& \frac{m^2}{H^2}
\frac{1}{(1+r)^2}
\sim \frac{m^2}{10^{-4}\times  \phi^2}\frac{1}{g_{PR}^6h^4}
<1,
\end{eqnarray}
which lead to a simple slow-roll condition $\phi_I > 10^{2} 
\times m/g_{PR}^{3}h^2$, where the field motion is {\bf not}
 oscillationary but simply slow-rolling.
Therefore, the usual requirement for instant preheating followed by the
instant decay now leads to a new condition for the model, which 
{\bf affects the initial condition of the oscillation} that determines
the amplitude of the oscillation;
dissipation leads to a delay of the end of chaotic
inflation. 
In fact, in conventional scenario of preheating after chaotic inflation,
the amplitude was simply assumed that $\phi\sim M_p \gg \phi_I$.
In contrast to the usual assumption, typical interaction of the scenario
may lead to dissipative motion that may lead to significant slow-roll
during $M_p> \phi>\phi_I$.
\begin{figure}[h]
 \begin{center}
\begin{picture}(0,120)(100,160)
\resizebox{8cm}{!}{\includegraphics{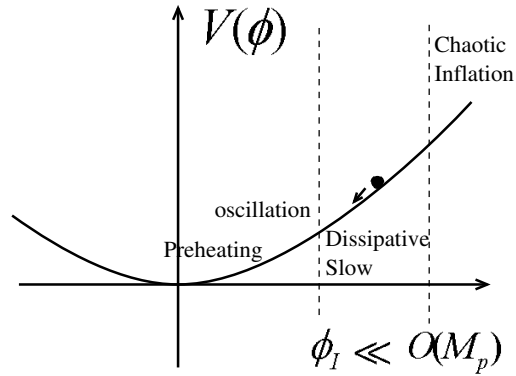}} 
\end{picture}
\caption{Without dissipation, chaotic inflation is expected to 
 end at $\phi\sim O(M_p)$.
However, interactions required for instant preheating suggests that 
non-thermal dissipation is common in such models and that an error
  may not be negligible.} 
 \end{center}
\end{figure}

\subsection{Remote inflation (Thermal inflation sourced continuously by dissipation)}
If dissipation is generic for many cosmological fields that acquire
$O(H)$ mass during inflation, these fields may lead to thermal
background during inflation.
Note that there are at least {\bf two} significant models in which
thermal background is used for inflationary scenario: warm inflation and
thermal inflation.
In fact, warm inflation is based on the thermal background sourced by
the dissipation, while no source mechanism has
been considered for thermal inflation.
Our idea is based on a naive question:
{\bf What happens if radiation in thermal inflation is sourced
continuously by (many) cosmological fields that acquire
$O(H)$ mass and dissipate their energy during inflation?}
Namely, background radiation sourced by dissipation may cause symmetry
restoration in a remote sector \cite{remote-inf, remote-inf2}, where
thermal inflation occurs. See Fig 4.
\begin{figure}[h]
 \begin{center}
\begin{picture}(0,170)(100,110)
\resizebox{9cm}{!}{\includegraphics{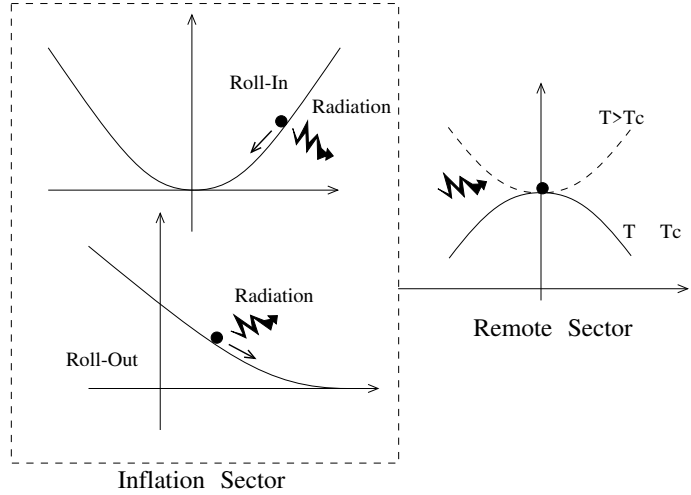}} 
\end{picture}
\caption{Left : Background radiation during inflation 
may be sourced by many dissipating fields.
Right: Radiation causes symmetry restoration in the remote sector, which
  leads to thermal inflation sourced by dissipation in the inflation
  sector. This model looks like a hybrid version of warm inflation.} 
 \end{center}
\end{figure}
There are two scenarios depending in which sector the slow-roll
condition is broken first.
(1) If the end of slow-roll occurs first in the inflaton sector, small
thermal inflation will remain in the remote sector.
Note that the temperature at this time is still higher than the critical
temperature of the thermal inflation sector.
(2) If the temperature decreases during inflation and 
symmetry breaking occurs first in the remote sector, it immediately
leads to the end of inflation. See Fig.4.

\subsection{Conclusions and discussions}
In this talk we considered simple examples of cosmological
scenarios in which dissipation may change the usual argument
based on cold (non-dissipating) scenario.
I believe the situation is now obvious suggesting why the dissipation
is very important for particle cosmology.
The study related to the cosmological dissipation may give us a key to
understand the interactions in the particle model
 in terms of the cosmological observations.
In previous studies dissipation has been studied only for the 
inflaton in the very early Universe (warm inflation).
However, in the light of particle cosmology, dissipation is
very important in understanding interactions in the SUSY
model \cite{D-moduli},  GUT or
the string theory \cite{Modulated, curvatonM, hybrid-kk, AD-brane}.

\end{document}